\documentclass[reprint,aps,prl,raggedbottom,twocolumn,superscriptaddress]{revtex4-2}
\usepackage{graphicx}
\usepackage{xifthen}
\usepackage{xspace}
\usepackage{amsmath}
\usepackage{amssymb}
\usepackage{comment}
\usepackage[colorlinks, allcolors=blue]{hyperref}
\usepackage[all]{hypcap}
\usepackage[mathlines]{lineno}
\usepackage{braket}
\usepackage{wrapfig}
\usepackage{lipsum}
\usepackage{ulem}
\usepackage{units}
\usepackage{xcolor}
\usepackage[utf8]{inputenc}
\usepackage{natbib}

\begin{document}

\title{Self‑Aligned Heterogeneous Quantum Photonic Integration}
\author{Kinfung Ngan}
\affiliation{
JILA and Department of Physics, University of Colorado Boulder, Colorado 80309, USA
}
\author{Yeeun Choi}
\affiliation{
Center for Quantum Technology, Korean Institute of Science and Technology (KIST), Seoul 02792, Republic of Korea
}
\affiliation{KU-KIST Graduate School of Converging Science and Technology, Korea University, Seoul, Republic of Korea}
\author{Chun-Chieh Chang}
\affiliation{
Center for Integrated Nanotechnologies, Los Alamos National Laboratory, Los Alamos, New Mexico 87545, USA
}
\author{Dongyeon Daniel Kang}
\email{dykang@kist.re.kr}
\affiliation{
Center for Quantum Technology, Korean Institute of Science and Technology (KIST), Seoul 02792, Republic of Korea
}
\affiliation{
Division of Quantum Information, KIST School, Korea University of Science and Technology (UST), Seoul 02792, Republic of Korea
}
\author{Shuo Sun}
\email{shuosun@colorado.edu}
\affiliation{
JILA and Department of Physics, University of Colorado Boulder, Colorado 80309, USA
}

\begin{abstract}
Integrated quantum photonics holds significant promise for scalable photonic quantum information processing, quantum repeaters, and quantum networks, but its development is hindered by the mismatch between materials hosting high-quality quantum emitters and those compatible with mature photonic technologies. Heterogeneous integration offers a potential solution to this challenge, yet practical implementations have been limited by inevitable insertion losses at material interfaces. Here, we present a self-aligned heterogeneous quantum photonic integration approach that can deterministically achieve near-unity coupling efficiency at the interface. To showcase our approach, we demonstrate Purcell enhancement of a silicon vacancy (SiV) center in diamond induced by a heterogeneous photonic crystal cavity defined by titanium dioxide (TiO$_2$), as well as optical spin control and readout via a TiO$_2$ photonic circuit. We further show that, when combined with inverse photonic design, our approach enables efficient and broadband collection of single photons from a color center into a heterogeneous waveguide. Our approach is not restricted to SiV centers or TiO$_2$; it can be broadly applied to integrate diverse solid-state quantum emitters with thin-film photonic devices where conformal deposition is possible. Together, these results establish a practical route to scalable quantum photonic integrated circuits that combine high-quality quantum emitters with technologically mature photonic platforms.

\end{abstract}

\maketitle

\section{Introduction}
A long-standing goal in quantum photonics is the development of an integrated photonics platform that truly operates in the quantum regime~\cite{atature2018material,wang2020integrated}. If such a platform can achieve sufficiently high fidelity, efficiency, and scalability, it could enable transformative applications ranging from optical quantum computing~\cite{kok2007linear} and quantum optical neural networks~\cite{basani2025universal}, to quantum repeaters and quantum networks~\cite{azuma2023quantum}. What distinguishes quantum integrated photonics from its classical counterpart is its ability to generate, process, and store quantum states of light. Realizing this capability, however, is challenging because it requires strong optical nonlinearities at the single-photon level~\cite{chang2014quantum}. 

Solid-state quantum emitters, such as semiconductor quantum dots, color centers, and organic molecules, have demonstrated remarkable performance as deterministic sources of single~\cite{aharonovich2016solid,senellart2017high,toninelli2021single} and entangled photons~\cite{huber2018semiconductor}. Moreover, many of these emitters host internal electron and nuclear spin states that can be harnessed for storing photonic quantum states~\cite{nguyen2019quantum}, mediating photon-photon interactions~\cite{sun2018single,bhaskar2020experimental}, and generating more complex photonic cluster states~\cite{buterakos2017deterministic,zhan2020deterministic}. Integrating such quantum emitters into photonic integrated circuits holds tremendous promise for advancing photonic quantum computing~\cite{kok2007linear,basani2025universal} and quantum networking~\cite{azuma2023quantum}. Despite this potential, progress has been hindered by the disparate and often incompatible material requirements of the individual components. For example, group-IV color centers in diamond are among the leading candidates for realizing spin–photon interfaces~\cite{bradac2019quantum,janitz2020cavity}. However, despite advances in diamond nanofabrication~\cite{burek2012free,khanaliloo2015high,wan2018two,ding2024high}, building large-scale photonic circuits directly in diamond remains a grand challenge. Conversely, materials such as silicon~\cite{siew2021review}, silicon nitride~\cite{xiang2022silicon}, and lithium niobate~\cite{zhu2021integrated} offer excellent scalability for photonic integration, but there are no quantum emitters identified in these materials that possess both optical and spin properties comparable to those available in diamond.

\begin{figure*}[t!]
     \centering
         \includegraphics[width=1\textwidth]{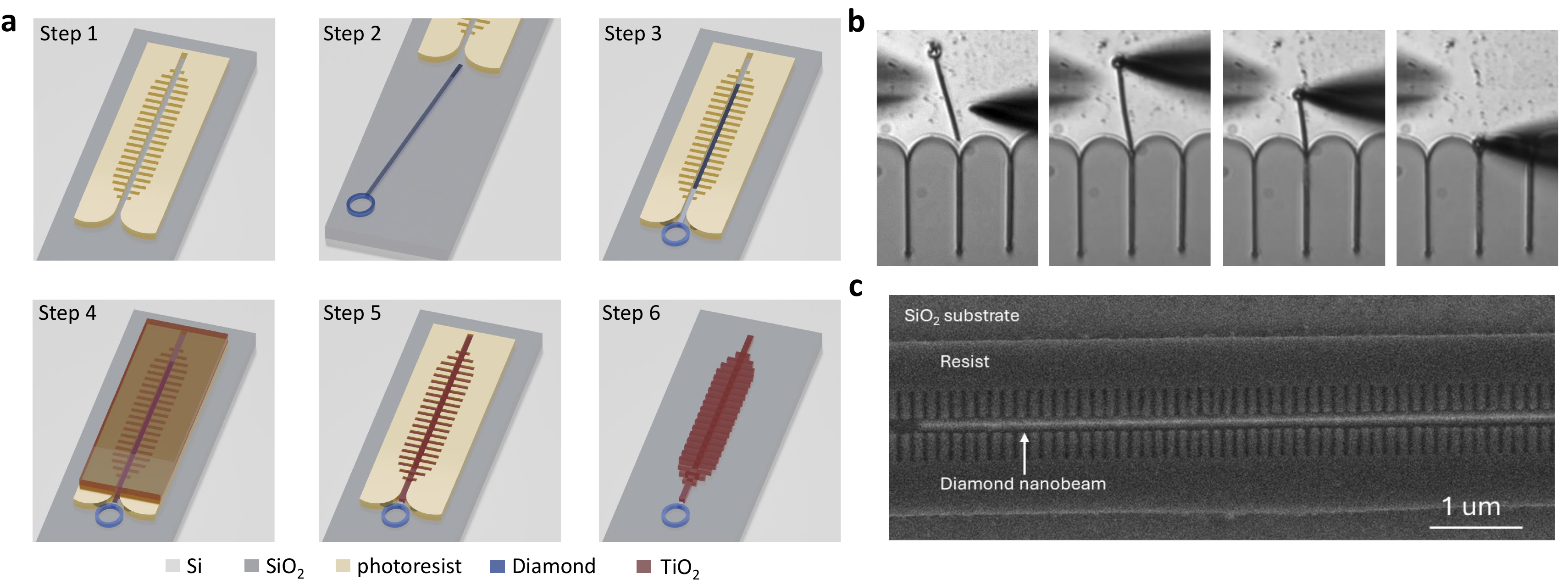}
         \caption{Self-aligned heterogeneous photonic integration. (a) Workflow of the self-aligned heterogeneous photonic integration. Step 1: patterning of the inverse of the full photonic circuit into a photoresist layer. Step 2: pick-and-place of a diamond nanobeam near the entrance of the slot defined by the photoresist. Step 3: insertion of the diamond nanobeam into the slot via a self-guided process. Step 4: conformal deposition of TiO$_2$ over the entire substrate via atomic layer deposition. Step 5: removal of excess TiO$_2$ via back etching. Step 6: removal of the photoresist. (b) Optical microscope images showing different stages during the nanobeam insertion process. (c) SEM image of the device right after the insertion of the diamond nanobeam (Step 3).}
         \label{fig1}
\end{figure*}

Heterogeneous integration offers a promising path by combining disparate materials within a single platform, thereby leveraging the advantages of both~\cite{elshaari2020hybrid,kim2020hybrid}. This approach has been highly successful in classical photonics, for example in integrating III–V lasers with silicon~\cite{liang2010recent} or other low-loss materials~\cite{xiang2021high,nader2025heterogeneous}. In quantum photonics, however, heterogeneous integration presents unique challenges, as even modest insertion loss at material interfaces can destroy the non-classical features of quantum light. At present, no strategy achieves both low loss and broad material compatibility. Direct embedding of quantum nanoparticles into photonic devices made in another material often introduces substantial scattering and intra-cavity loss~\cite{sahoo2022hybrid,ngan2023quantum,bayer2023optical,lettner2024controlling}. Adiabatic mode transfer based on pick-and-place~\cite{mouradian2015scalable,kim2017hybrid,chanana2022ultra,riedel2023efficient}, transfer-printing~\cite{wan2020large,takada2025alignment}, and lock-and-release~\cite{li2024heterogeneous} techniques typically introduce dB-level insertion loss due to alignment inaccuracies. Top-down heterogeneous integration~\cite{davanco2017heterogeneous} improves alignment accuracy, yet its applicability is restricted to host materials that can be fabricated as large membranes and reliably bonded to a heterogeneous substrate, making systems such as diamond particularly challenging. Overcoming these losses and alignment constraints remains essential for realizing scalable quantum photonic integrated circuits.

In this Article, we present a heterogeneous quantum photonic integration approach that can achieve near-unity coupling efficiency at material interfaces through fully self-aligned components. We showcase the method by integrating diamond with titanium dioxide (TiO$_2$), a material recently identified as promising for scalable integrated photonics~\cite{butcher2020high, chen2025development}. Using the resulting devices, we demonstrate Purcell enhancement of a silicon-vacancy (SiV) center in diamond induced by a diamond-TiO$_2$ heterogeneous photonic crystal cavity, as well as optical spin control and readout of a SiV center via a TiO$_2$ photonic circuit. We further demonstrate that our approach, when combined with inverse photonic design, could enable efficient and broadband collection of single photons from a quantum emitter into a heterogeneous waveguide. Our approach is not limited to color centers in diamond or TiO$_2$ photonics. It is broadly applicable to virtually any solid-state quantum emitter integrated with thin-film photonic devices where conformal deposition is feasible. Together, these results represent a significant step toward scalable, high-efficiency integration of solid-state quantum emitters with large-scale photonic circuits.

\section{Results}
\subsection{Workflow of the Self-Aligned Heterogeneous Photonic Integration}

Figure~\ref{fig1}a outlines the workflow of the self-aligned heterogeneous photonic integration. We illustrate the process using a heterogeneous fishbone photonic crystal cavity. However, as we will discuss later, the scheme is general and can be applied to a wide range of photonic geometries. First, we pattern the inverse of the full photonic circuit into a photoresist layer on a SiO$_2$ substrate (Step 1). A diamond nanobeam is then inserted into its designated slot defined by the photoresist (Steps 2 and 3). Following insertion, we conformally deposit TiO$_2$ across the chip (Step 4), remove the excess TiO$_2$ by back-etching (Step 5), and strip the resist (Step 6). This sequence transfers the resist pattern into TiO$_2$ and leaves the diamond nanobeam seamlessly embedded within the TiO$_2$ photonic circuit. The \textbf{Materials and methods} section contains details of the heterogeneous device fabrication.

The insertion of the diamond nanobeam is the most critical step in our heterogeneous photonic integration. The diamond nanobeams are fabricated on an electronic-grade single-crystal diamond via electron beam lithography followed by angled etching in a Faraday cage (see \textbf{Materials and methods}). Using a tungsten probe under an optical microscope, we first pick up a nanobeam from the diamond substrate and transfer it to the heterogeneous substrate near the designated slot. We then use the same probe to push the nanobeam into the slot. A ring feature at one end of the nanobeam facilitates engagement between the probe and the beam. To achieve self-alignment, the slot width (230~nm) is designed to be marginally larger than the width of the nanobeam (200~nm). Combined with the large aspect ratio of the nanobeam, this geometry constrains the lateral and angular degrees of freedom and promotes straight insertion. Because the slot and the nanobeam are both subwavelength in width, they are not resolvable under the optical microscope. To address this, the resist pattern incorporates a funnel-shaped entrance that passively guides the nanobeam. As the probe advances the beam, the funnel progressively corrects its orientation, and once engaged, the beam naturally registers with the slot. Figure~\ref{fig1}b shows several microscope images captured during the insertion. Figure~\ref{fig1}c shows a scanning electron microscopy (SEM) image after successful insertion, which confirms a straight, well-seated nanobeam with no observable damage to the fishbone photonic crystal pattern defined by the resist.

\begin{figure}[t!]
     \centering
         \includegraphics[width=0.5\textwidth]{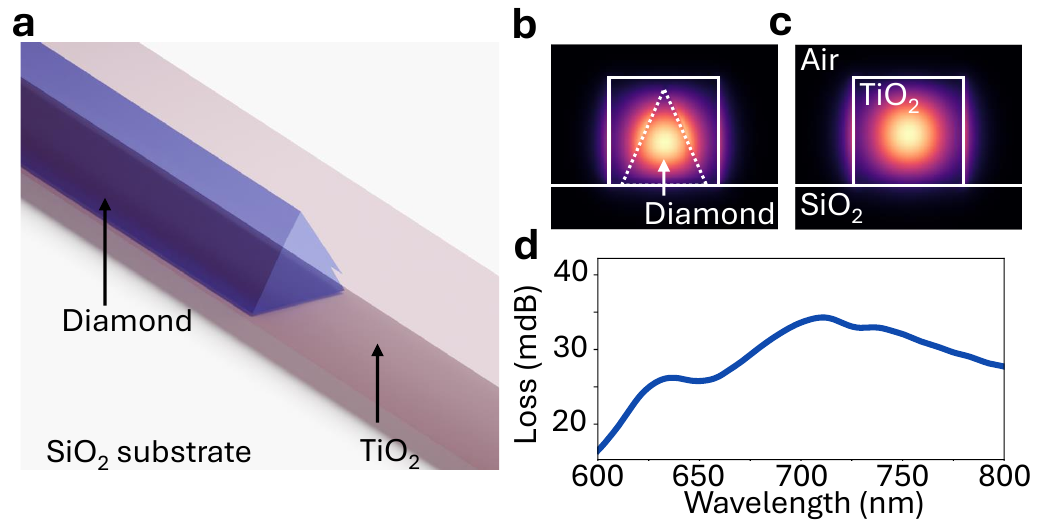}
         \caption{Analysis of external coupling loss for our self-aligned heterogeneous photonic devices. (a) Schematic illustration of the interface at the diamond-TiO$_2$ heterogeneous waveguide and a monolithic TiO$_2$ waveguide. (b), (c) Calculated electric-field intensity profile in the cross-section plane of the diamond-TiO$_2$ heterogeneous and TiO$_2$ monolithic waveguides. (d) Calculated insertion loss at the interface between the two waveguides.}
         \label{fig2}
\end{figure}

The primary advantage of our heterogeneous integration approach is that it introduces minimal intracavity and out-coupling losses. Because the diamond nanobeam extends across the entire cavity region, the cavity mode is fully supported by the hybrid diamond/TiO$_2$ structure. Consequently, the heterogeneous interface does not, in principle, induce any intracavity loss, as corroborated by the high quality factors obtained from both simulations and experimental measurements, which will be discussed in the following section. To evaluate the out-coupling efficiency, we consider the interface between a hybrid diamond/TiO$_2$ ridge waveguide and a monolithic TiO$_2$ ridge waveguide of identical width, as illustrated in Fig.~\ref{fig2}a. Figures~\ref{fig2}b and \ref{fig2}c show the electric field intensity distributions of the fundamental modes in the heterogeneous and monolithic waveguides, respectively. The two modes exhibit nearly identical spatial profiles and effective refractive indices of 1.89 and 1.79 at 737~nm, owing to the closely matched refractive indices of diamond and TiO$_2$. Figure~\ref{fig2}d shows the simulated insertion loss at the butt-coupled interface. The device achieves an insertion loss of less than 0.034 dB ($<0.8\%$) over a broad wavelength range. This insertion loss can be further reduced by tapering the diamond nanobeam for adiabatic mode conversion.

\subsection{Purcell Enhancement by a Heterogeneous Photonic Cavity}

\begin{figure*}[t!]
     \centering
       \includegraphics[width=1\textwidth]{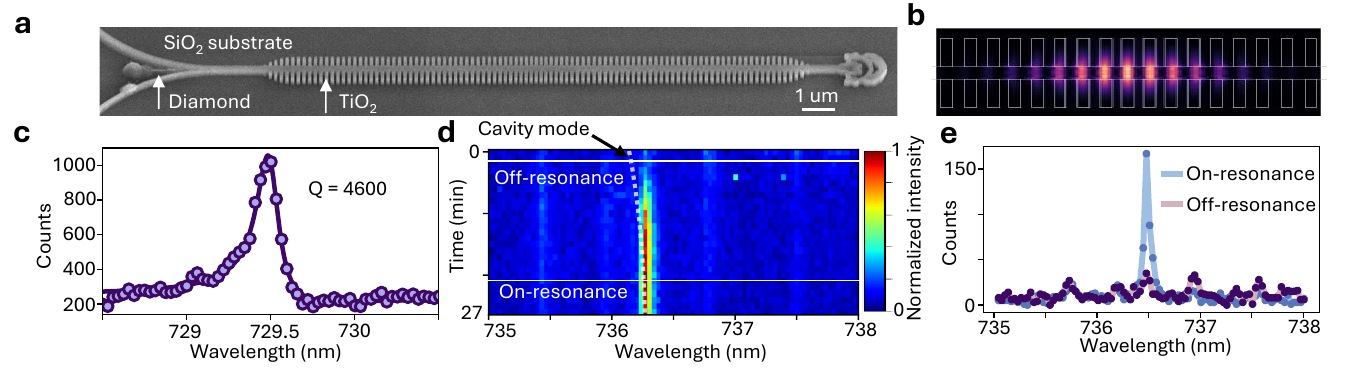}
         \caption{Purcell enhancement induced by a heterogeneous photonic crystal cavity. (a) SEM image of the diamond-TiO$_2$ heterogeneous photonic crystal cavity. (b) Calculated longitudinal electric field intensity profile of the heterogeneous photonic cavity. (c) Transmission spectrum of a heterogeneous photonic crystal cavity, yielding a $Q$ factor of 4600(180). (d) Photoluminescence spectra of an embedded SiV center as the cavity resonance is tuned in time via gas condensation. The white dashed line indicates the resonant frequency of the cavity mode. (e) Photoluminescence spectra of a SiV center when it is resonant (blue) and far detuned (red) from the cavity mode. For (c) - (e), the spectra was obtained by exciting the cavity or the SiV center via free space and collecting the signal from the inverse-designed grating coupler on the right end.}
         \label{fig3}
\end{figure*}

We first demonstrate a fishbone photonic crystal cavity based on our self-aligned heterogeneous integration approach. Figure~\ref{fig3}a shows the SEM image of the heterogeneous diamond-TiO$_2$ fishbone photonic crystal cavity. The cavity design is based on a diamond nanobeam with a triangular cross-section embedded inside a TiO$_2$ fishbone photonic crystal. Figure~\ref{fig3}b shows the simulated electric field intensity of the cavity mode in the propagation plane. The cavity features a field maximum in the center of the diamond nanobeam, crucial for maximizing the coupling with embedded color centers. The cavity resonance is designed to be $\lambda = 737$~nm, matching the zero-phonon-line of the SiV center. It has a simulated $Q$ factor of 120,000 and a mode volume of 1.52$(\frac{\lambda}{n})^{3}$, where $n=2.216$ is the refractive index of TiO$_2$ at 737~nm. 

Figure~\ref{fig3}c shows the experimentally measured transmission spectrum of the fishbone photonic crystal cavity (see \textbf{Materials and methods} for the experimental setup used throughout this work). The cavity transmission is measured by exciting the cavity center from free space with a super-continuum laser, and collecting the transmitted laser intensity via the inverse-designed grating coupler at the end of the waveguide (see \textbf{Supplementary Section 1} for the design and characterization of the inverse-designed grating coupler). By fitting the measured spectrum (circles) to a Fano lineshape (solid line), we deduce a cavity linewidth of 0.16~nm and a cavity $Q$ factor of 4,600. The calculated cavity $Q$ factor based on the geometry extracted from the SEM image is 5,200 (see \textbf{Supplementary Section 2}), in good agreement with the experimental result. This agreement suggests that the cavity $Q$ factor is primarily limited by geometric deviations between the design and the fabricated device, rather than by additional scattering or absorption introduced by the heterogeneous integration or other steps in our fabrication process.

To examine the Purcell enhancement induced by the heterogeneous photonic crystal cavity, we identify a different device in which a SiV center lies near the cavity center. We measure the SiV photoluminescence by exciting it with a 532-nm laser from free space and collecting the emission through the same inverse-designed grating coupler at the end of the waveguide. Figure~\ref{fig3}d shows the SiV photoluminescence as we tune the cavity wavelength via gas condensation. We observe a strong increase in photoluminescence intensity as the cavity is tuned into resonance with the SiV center. Figure~\ref{fig3}e compares the photoluminescence spectra when the cavity is far detuned from (red) and on resonance with (blue) the SiV center. From the intensity enhancement, we deduce a minimum Purcell factor of 6. The achieved Purcell factor is limited because the SiV center is not located exactly at a field antinode of the cavity mode, and because the specific cavity used in this measurement has a lower $Q$ of only 640 (see \textbf{Supplementary Section 3}). Nonetheless, these measurements demonstrate Purcell enhancement in our heterogeneous photonic cavity. With an improved $Q$ factor (4,600) and SiV position, a Purcell factor of 150 can be achieved.

\subsection{Chip-Integrated Optical Spin Control and Readout}

\begin{figure*}[t!]
     \centering
         \includegraphics[width=\textwidth]{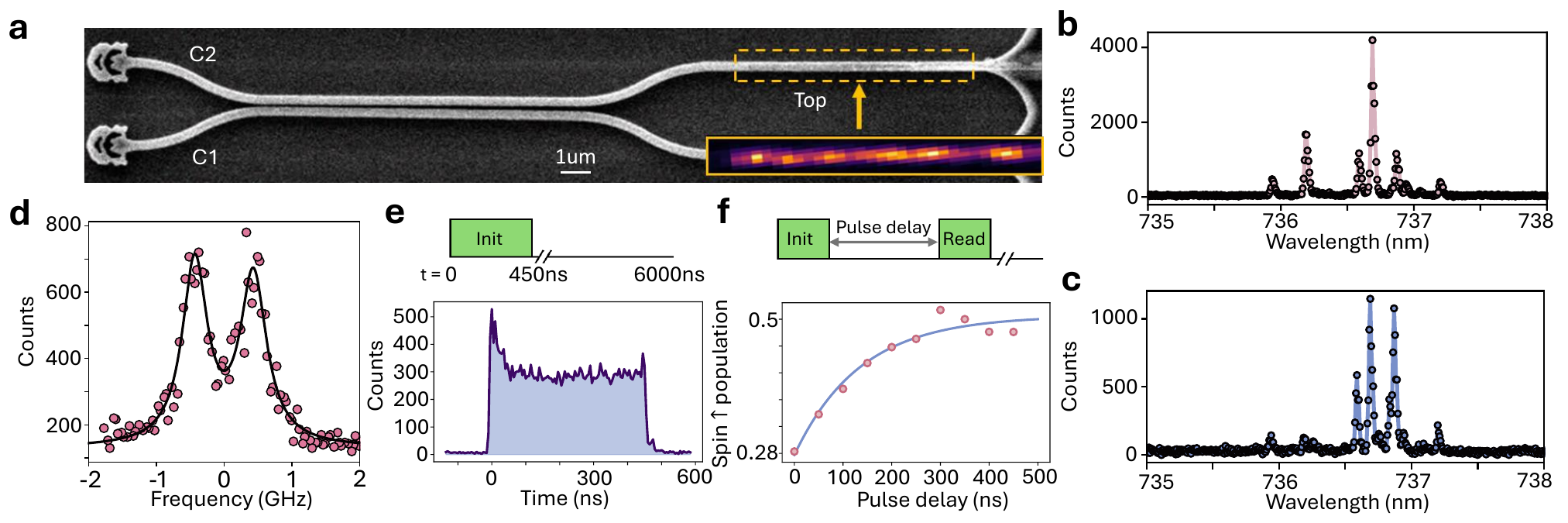}
         \caption{Optical spin initialization and readout via a TiO$_2$ photonic circuit. (a) SEM image of a TiO$_2$ 80/20 insertion coupler integrated with a diamond nanobeam. The diamond nanobeam is embedded inside the upper arm of the TiO$_2$ waveguide. The inset shows the scanning confocal photoluminescence image of the region with the inserted diamond nanobeam, showing a large density of SiV centers embedded inside. (b) Photoluminescence spectra of the SiV centers collected from the top of the integrated diamond nanobeam via free space. (c) Photoluminescence spectra of the same SiV centers collected from the inverse-designed grating coupler at port C1. (d) Photoluminescence excitation spectrum of the optical C transition of a specific SiV center under a magnetic field around 0.3 T. The two peaks correspond to the two spin-conserving transitions. (e) Optical initialization and readout of the SiV spin via the photonic circuit. (f) Measurement of spin relaxation time via a pump-probe sequence.}
         \label{fig4}
\end{figure*}

Besides cavities, our heterogeneous integration approach enables the direct incorporation of diamond color centers into virtually any TiO$_2$ photonic circuits. As an example, we demonstrate the integration of diamond SiV centers with a 2×2 TiO$_2$ insertion coupler. Figure~\ref{fig4}a shows the SEM image of the TiO$_2$ insertion coupler integrated with a diamond nanobeam from the top right channel. The diamond nanobeam contains a large density of SiV centers, as demonstrated by the scanning photoluminescence image of the region around the inserted diamond nanobeam (inset of Fig.~\ref{fig4}a). To demonstrate coupling between the embedded SiV centers and the TiO$_2$ photonic circuit, we excite the center of the diamond nanobeam with a 532-nm laser from free space, and collect the photoluminescence spectra from both the top (Fig.~\ref{fig4}b) and one of the inverse-designed grating couplers following the insertion coupler (Fig.~\ref{fig4}c). The spectra are nearly identical between the two collection channels, confirming that the SiV emission is efficiently coupled into the TiO$_2$ photonic circuit. We attribute the difference in the relative intensities of the emission peaks between Fig.~\ref{fig4}b and Fig.~\ref{fig4}c to the polarization-dependent coupling between the SiV emission and the heterogeneous waveguide.

Building on this SiV-integrated TiO$_2$ platform, we further demonstrate optical control and readout of a SiV spin through the TiO$_2$ photonic circuit. Such chip-integrated optical spin control and readout are essential for multiplexing many quantum memories on a single chip, which is a key requirement for scalable quantum repeaters and quantum networks~\cite{munro2010quantum}. We identify a specific SiV center based on its zero-phonon-line wavelength and apply an external magnetic field of 0.3 T to lift the spin degeneracy. Figure~\ref{fig4}d shows the photoluminescence excitation spectrum of the optical C transition from this SiV center, revealing two spin-conserving transitions separated by 1 GHz. To demonstrate optical spin control and readout, we inject a resonant laser into the chip via the inverse-designed grating coupler C2 and selectively drive one of the spin-conserving transitions, while collecting the phonon-sideband emission from the coupler C1. The resonant excitation is modulated by an acousto-optic modulator (AOM) into a periodic 450-ns pulse train with a 6-$\mu$s repetition period, as shown in the top panel of Fig.~\ref{fig4}e. The bottom panel of Fig.~\ref{fig4}e displays the time-resolved histogram of the phonon-sideband signal, which exhibits a clear exponential decay characteristic of optical pumping. These measurements demonstrate that both optical spin initialization and readout can be performed directly through the TiO$_2$ photonic circuit.

From the optical pumping curve, we estimate an initialization fidelity of 72$\%$. This relatively modest fidelity is likely limited by spin relaxation during the initialization sequence, since the spin relaxation time $T_1$ is comparable to the optical initialization time. Figure~\ref{fig4}f shows a pump–probe measurement of the spin relaxation time $T_1$, yielding 130~ns. From an exponential fit to the optical pumping dynamics, we extract an optical initialization time of 25~ns. We expect that we can improve the spin lifetime by purposely aligning the magnetic field with the SiV symmetry axis~\cite{sukachev2017silicon}, and speed up the optical initialization by driving a spin–non-conserving transition that provides a faster optical decay pathway.

\subsection{One-Way Quantum Light Emission via Inverse Design}

\begin{figure}[t!]
     \centering
         \includegraphics[width=0.5\textwidth]{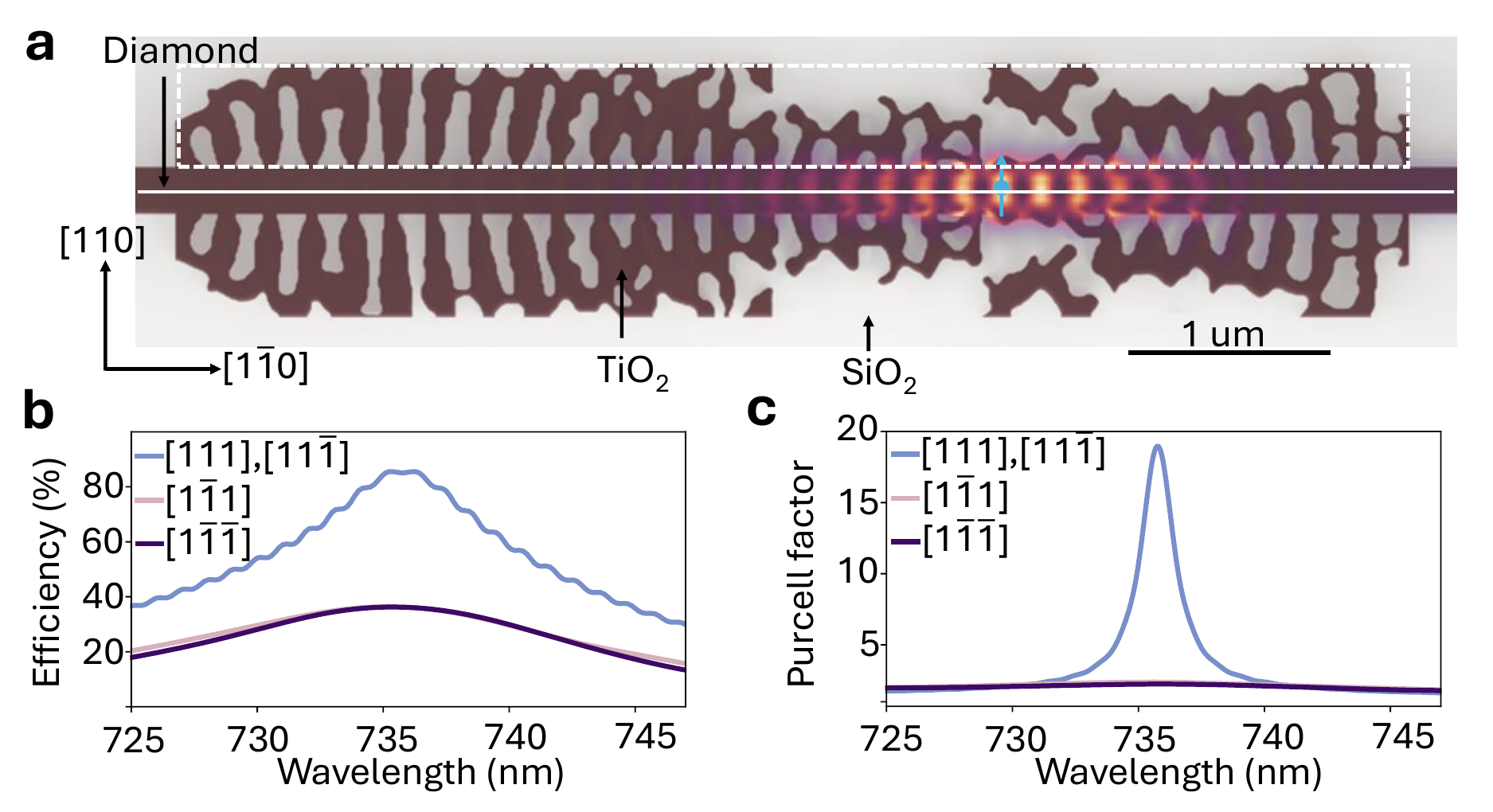}
         \caption{An inverse-designed one-way quantum light extractor. (a) Design geometry of the quantum light extractor overlaid with the calculated electric field distribution when excited by a dipole (blue arrow) polarized along the TE-mode of the waveguide. (b),(c) Calculated collection efficiency (b) and Purcell factor (c) for four possible orientations of a SiV center.}
         \label{Fig5}
\end{figure}

One unique advantage of our heterogeneous photonic platform is its full compatibility with inverse-designed photonic devices~\cite{molesky2018inverse}, enabling highly compact and broadband quantum photonic components. This is in stark contrast to heterogeneous integrated photonic circuits based on adiabatic mode conversion. To showcase this capability, we employ inverse photonic design to realize a compact, efficient, and broadband quantum light extractor that collects dipole emission from a single color center into a single waveguide mode propagating only in one direction. Such devices are essential for chip-integrated single-photon sources~\cite{aharonovich2016solid,senellart2017high,toninelli2021single} and reflection-based spin-photon interactions~\cite{nguyen2019quantum,sun2018single,bhaskar2020experimental}. Figure~\ref{Fig5}a shows the resulting design, overlaid with the electric field distribution when excited by an ideal dipole source polarized along the TE mode of the waveguide. The dipole radiation is efficiently converted into the fundamental TE mode of the heterogeneous waveguide, as indicated by the electric-field mode profile. Note that our heterogeneous photonic platform imposes additional fabrication constraints because the region reserved for diamond nanobeam insertion cannot be etched. To account for this constraint, we define the design region entirely outside the insertion area, as indicated by the white dashed box in Fig.~\ref{Fig5}a, and we impose mirror symmetry about the waveguide centerline (white solid line). This approach preserves full patterning freedom in the TiO$_2$ layer while ensuring that the diamond insertion region remains unetched.

To quantitatively evaluate the performance of the quantum light extractor for a SiV center, we consider a diamond nanobeam fabricated from [001]-cut single-crystal diamond, with the TE mode of the waveguide polarized along the [110] direction. We assume an ideal SiV center whose optical C transition dipole is aligned with its symmetry axis~\cite{rogers2014electronic}. This symmetry axis, however, has four possible orientations. Figures~\ref{Fig5}b and~\ref{Fig5}c show the calculated collection efficiency and Purcell factor for all four orientations of the SiV center when it is used as the dipole source. The quantum light extractor achieves a peak collection efficiency of 84\% and a peak Purcell factor of 18 for the nearly-TE-polarized ($[111]$, $[11\bar{1}]$) orientations. Importantly, in contrast to conventional cavity-based designs, the inverse-designed device maintains a collection efficiency above 70\% across a 5.5-nm wide spectral window. We attribute this broadband performance to the fact that quantum light extraction in this architecture is fundamentally a mode-conversion process: efficient conversion from a localized dipole field to a single guided mode can be achieved over a wide spectral range, despite that the local density of states enhancement is more wavelength selective (Fig.~\ref{Fig5}c). For the nearly-TM-polarized ($[1\bar{1}1]$, $[1\bar{1}\bar{1}]$) orientations, the collection efficiency remains low at around 30\% with almost no Purcell enhancement, which is consistent with the design being optimized for a TE-polarized dipole source.

\section{Discussion}
The self-aligned heterogeneous integration strategy demonstrated here provides a practical pathway for integrating high-quality solid-state quantum emitters with thin-film photonic technologies. By embedding single-crystal diamond nanobeams directly into TiO$_2$ photonic circuits with near-perfect mode matching, our approach eliminates the interface losses and alignment tolerances that have long hindered heterogeneous quantum photonics. This architecture preserves the optical performance of bulk-crystal emitters while leveraging the scalability and design flexibility of thin-film photonic platforms.

In this work, we employ Atomic Layer Deposition (ALD) for conformal encapsulation of the inserted diamond nanobeam. Although ALD offers excellent conformality, it is inherently slow and not ideal for producing photonic thin films with thicknesses of several hundred nanometers. Importantly, the underlying integration concept is compatible with other deposition methods that can be engineered to produce conformal thin-film coatings, such as Physical Vapor Deposition (PVD)~\cite{alfaro2021highly} or Plasma-Enhanced Chemical Vapor Deposition (PECVD)~\cite{peter2010low}. Developing conformal deposition processes based on these technologies would provide significantly higher throughput and access to a broader class of photonic materials.

Our platform creates promising opportunities for advancing quantum photonic technologies. With improved cavity fabrication and deterministic emitter placement, cavity $Q$ factors above 10,000 and Purcell factors exceeding 300 should be attainable, enabling high cooperativity for spin-photon interfaces~\cite{aharonovich2016solid,senellart2017high,toninelli2021single}, quantum optical nonlinearities~\cite{sun2018single,bhaskar2020experimental}, and photonic quantum registers~\cite{nguyen2019quantum}. Furthermore, the capability to integrate many quantum emitters within large-scale photonic circuits, combined with inverse-designed photonic devices, makes it possible to realize complex on-chip networks featuring wavelength-selective routing, spatial-mode engineering, and multiplexed spin control and readout, capabilities essential for quantum repeaters based on multiplexed quantum memories~\cite{munro2010quantum}.

Beyond circuit-level applications, the heterogeneous photonic architecture naturally supports exploration of photon-mediated collective and many-body interactions by embedding multiple emitters within an engineered photonic environment~\cite{chang2018colloquium}. Such systems could enable experimental access to dissipative many-body dynamics~\cite{masson2020many}, long-range correlations~\cite{gonzalez2015subwavelength}, and collective entanglement generation~\cite{aron2016photon}, opening a pathway toward creating synthetic quantum materials based on coupled quantum emitters.

\section{Materials and methods}

\subsection{Fabrication of the heterogeneous photonic device}
We start the device fabrication by spin-coating a 320-nm-thick layer of ZEP520A onto an HMDS-primed SiO$_2$ substrate. The HMDS layer improves the adhesion and mechanical stability of the patterned resist during the subsequent diamond insertion. We then define the photonic structure by exposing the resist with the inverse pattern of the device using an electron-beam writer, followed by resist development. After development, we perform 3 s of O$_2$ plasma ashing to remove residual developed resist and exposed HMDS, while leaving the unexposed resist nearly intact. Next, we apply deionized water to the SiO$_2$ surface to reduce adhesion between the nanobeam and the substrate, followed by insertion of the diamond nanobeam as described in Fig.~\ref{fig1}. After insertion, we perform conformal deposition of TiO$_2$ by ALD at $90^{\circ}\mathrm{C}$ to a thickness sufficient to fully embed the inserted nanobeam and fill the slot. We then etch back the overgrown TiO$_2$ until the resist is re-exposed. Finally, we remove the resist using PG Remover at $70^{\circ}\mathrm{C}$ for 20 min, followed by a 3-min Nanostrip clean at $80^{\circ}\mathrm{C}$ to remove etch residues, and anneal the device on a hotplate at $250^{\circ}\mathrm{C}$ for 2 h to reduce material absorption loss.

\subsection{Fabrication of the diamond nanobeam}

Diamond fabrication started with silicon ion implantation into electronic-grade diamond at 190 keV with a fluence of $10^{12}$ ions per cm$^2$. High-temperature annealing was then performed at 1150$^{\circ}\mathrm{C}$ under high vacuum ($10^{-6}$ Torr) to activate optically active defects. Diamond nanobeams were fabricated using an angled-etching technique with a PECVD-deposited silicon nitride hard mask with a thickness of 300~nm. The hard mask was patterned using electron-beam lithography and fluorine-based reactive ion etching to define the nanobeam geometry. Diamond etching was then carried out using an oxygen-based inductively coupled plasma process. The mask pattern was initially transferred into the diamond by a shallow anisotropic etch to establish the nanobeam profile, followed by angled etching using a custom-designed Faraday cage. The Faraday cage redirected incoming ions toward the substrate at oblique angles, enabling controlled undercutting and the formation of fully suspended diamond nanobeams with triangular cross-sections. This geometry was optimized to facilitate subsequent pick-up and transfer processes. After etching, the silicon nitride mask was removed by chemical treatments, yielding clean and well-defined diamond nanobeam structures.

\subsection{Measurement setup}
We mounted the sample in a closed-cycle cryostat (Montana Instruments) and cooled it to 5 K. Optical excitation and signal collection were carried out using a home-built confocal microscope with a vacuum-compatible objective lens with a numerical aperture of 0.9. For cavity transmission measurements (Fig.~\ref{fig3}c), we excited the cavity with a supercontinuum laser and collected the transmitted signal with a spectrometer (Andor). For photoluminescence spectroscopy (Fig.~\ref{fig3}d and \ref{fig3}e, Fig.~\ref{fig4}b and \ref{fig4}c), we excited the SiV centers with a 532-nm diode laser (Edmund Optics) and collected the emission with the same spectrometer. For photoluminescence excitation spectroscopy (Fig.~\ref{fig4}d), we excited the SiV center near-resonantly with a tunable continuous-wave Ti:sapphire laser (M-Squared) and collected the phonon-sideband emission with a single-photon counting module (Excelitas). For optical spin control and readout (Fig.~\ref{fig4}e and \ref{fig4}f), we mounted the sample on top of a SmCo permanent magnet that applied a 0.3 T field to lift the spin degeneracy. We generated laser pulse sequences by sending the Ti:sapphire output through an acousto-optic modulator (AA Optoelectronic) and time-gating it with a digital signal generator (Swabian Instruments).

\subsubsection{\normalfont \bfseries Acknowledgments}

K.N. and S.S. acknowledge funding from the National Science Foundation (Grant No. 2317149 and 2443684), the W. M. Keck Foundation, and the Sloan Research Fellowship. Y.C. and D.D.K. acknowledge funding from the KIST institutional program (2E33541, 2E33571) funded by the Korea Institute of Science and Technology, the National Research  Foundation of Korea (NRF) grants (Nos. RS-2022-NR068818, RS-2025-25445839), and the Institute for Information $\&$ Communication Technology Planning $\&$ Evaluation (IITP) grant (No. RS-2025-25464657). The device fabrication was performed, in part, at the Center for Integrated Nanotechnologies, an Office of Science User Facility operated for the U.S. Department of Energy (DOE) Office of Science. Los Alamos National Laboratory, an affirmative action equal opportunity employer, is managed by Triad National Security, LLC for the U.S. Department of Energy’s NNSA, under contract 89233218CNA000001. The authors acknowledge Andrew Mounce, Yijun Xie, Seung-Woo Jeon, Ting-Wei Hsu, Yuan Zhan, Jakob M. Grzesik, and Hope Lee for helpful discussions.

\subsubsection{\normalfont \bfseries Author Contributions}

K.N. and S.S. conceived and designed the experiment. K.N. developed the heterogeneous integration method, designed and fabricated all the devices, set up and performed the measurements, and analyzed the data. Y.C. and D.D.K. fabricated the diamond nanobeam and performed thermal annealing for the activation of the color centers. C.C.C performed static cavity resonance shift via ALD. D.D.K. and S.S. supervised the whole project. All authors participated in the preparation of the manuscript.

\subsubsection{\normalfont \bfseries Data availability}
The data generated and/or analyzed during the current study will be available via a DOI link upon the publication of this paper.

\subsubsection{\normalfont \bfseries Conflict of interest}
The authors declare no competing interests.

\section*{References}

\bibliographystyle{naturemag}
\bibliography{ref}

@article{riedel2023efficient,
  title={Efficient photonic integration of diamond color centers and thin-film lithium niobate},
  author={Riedel, Daniel and Lee, Hope and Herrmann, Jason F and Grzesik, Jakob and Ansari, Vahid and Borit, Jean-Michel and Stokowski, Hubert S and Aghaeimeibodi, Shahriar and Lu, Haiyu and McQuade, Patrick J and others},
  journal={ACS Photonics},
  volume={10},
  number={12},
  pages={4236--4243},
  year={2023},
  publisher={ACS Publications}
}

@article{rogers2014electronic,
  title={Electronic structure of the negatively charged silicon-vacancy center in diamond},
  author={Rogers, Lachlan J and Jahnke, Kay D and Doherty, Marcus W and Dietrich, Andreas and McGuinness, Liam P and M{\"u}ller, Christoph and Teraji, Tokuyuki and Sumiya, Hitoshi and Isoya, Junichi and Manson, Neil B and others},
  journal={Physical Review B},
  volume={89},
  number={23},
  pages={235101},
  year={2014},
  publisher={APS}
}

@article{masson2020many,
  title={Many-body signatures of collective decay in atomic chains},
  author={Masson, Stuart J and Ferrier-Barbut, Igor and Orozco, Luis A and Browaeys, Antoine and Asenjo-Garcia, Ana},
  journal={Physical Review Letters},
  volume={125},
  number={26},
  pages={263601},
  year={2020},
  publisher={APS}
}

@article{gonzalez2015subwavelength,
  title={Subwavelength vacuum lattices and atom--atom interactions in two-dimensional photonic crystals},
  author={Gonz{\'a}lez-Tudela, Alejandro and Hung, C-L and Chang, Darrick E and Cirac, J Ignacio and Kimble, HJ},
  journal={Nature Photonics},
  volume={9},
  number={5},
  pages={320--325},
  year={2015},
  publisher={Nature Publishing Group UK London}
}

@article{aron2016photon,
  title={Photon-mediated interactions: A scalable tool to create and sustain entangled states of N atoms},
  author={Aron, Camille and Kulkarni, Manas and T{\"u}reci, Hakan E},
  journal={Physical Review X},
  volume={6},
  number={1},
  pages={011032},
  year={2016},
  publisher={APS}
}

@article{alfaro2021highly,
  title={Highly-conformal sputtered through-silicon vias with sharp superconducting transition},
  author={Alfaro-Barrantes, JA and Mastrangeli, M and Thoen, DJ and Visser, Sten and Bueno, J and Baselmans, JJA and Sarro, PM},
  journal={Journal of Microelectromechanical Systems},
  volume={30},
  number={2},
  pages={253--261},
  year={2021},
  publisher={IEEE}
}

@article{peter2010low,
  title={Low temperature plasma enhanced chemical vapor deposition of thin films combining mechanical stiffness, electrical insulation, and homogeneity in microcavities},
  author={Peter, S and G{\"u}nther, M and Hauschild, D and Richter, F},
  journal={Journal of Applied Physics},
  volume={108},
  number={4},
  year={2010},
  publisher={AIP Publishing}
}

@article{chang2018colloquium,
  title={Colloquium: Quantum matter built from nanoscopic lattices of atoms and photons},
  author={Chang, DE and Douglas, JS and Gonz{\'a}lez-Tudela, Alejandro and Hung, C-L and Kimble, HJ},
  journal={Reviews of Modern Physics},
  volume={90},
  number={3},
  pages={031002},
  year={2018},
  publisher={APS}
}

@article{munro2010quantum,
  title={From quantum multiplexing to high-performance quantum networking},
  author={Munro, WJ and Harrison, KA and Stephens, AM and Devitt, SJ and Nemoto, Kae},
  journal={Nature Photonics},
  volume={4},
  number={11},
  pages={792--796},
  year={2010},
  publisher={Nature Publishing Group UK London}
}

@article{molesky2018inverse,
  title={Inverse design in nanophotonics},
  author={Molesky, Sean and Lin, Zin and Piggott, Alexander Y and Jin, Weiliang and Vuckovi{\'c}, Jelena and Rodriguez, Alejandro W},
  journal={Nature Photonics},
  volume={12},
  number={11},
  pages={659--670},
  year={2018},
  publisher={Nature Publishing Group UK London}
}

@article{sukachev2017silicon,
  title={Silicon-vacancy spin qubit in diamond: a quantum memory exceeding 10 ms with single-shot state readout},
  author={Sukachev, Denis D and Sipahigil, Alp and Nguyen, Christian T and Bhaskar, Mihir K and Evans, Ruffin E and Jelezko, Fedor and Lukin, Mikhail D},
  journal={Physical Review Letters},
  volume={119},
  number={22},
  pages={223602},
  year={2017},
  publisher={APS}
}

@article{butcher2020high,
  title={High-Q nanophotonic resonators on diamond membranes using templated atomic layer deposition of TiO2},
  author={Butcher, Amy and Guo, Xinghan and Shreiner, Robert and Delegan, Nazar and Hao, Kai and Duda III, Peter J and Awschalom, David D and Heremans, F Joseph and High, Alexander A},
  journal={Nano Letters},
  volume={20},
  number={6},
  pages={4603--4609},
  year={2020},
  publisher={ACS Publications}
}

@article{chen2025development,
  title={Development of yellow-light TiO2 integrated photonics},
  author={Chen, Zequn and Tang, Yiheng and Wei, Maoliang and Li, Xiaojing and Sun, Boshu and Wu, Yingchun and Zou, Sishuo and Huang, Ji and Si, Ke and Gong, Wei and others},
  journal={Optics Letters},
  volume={50},
  number={11},
  pages={3489--3492},
  year={2025},
  publisher={Optica Publishing Group}
}

@article{davanco2017heterogeneous,
  title={Heterogeneous integration for on-chip quantum photonic circuits with single quantum dot devices},
  author={Davanco, Marcelo and Liu, Jin and Sapienza, Luca and Zhang, Chen-Zhao and De Miranda Cardoso, Jos{\'e} Vin{\'\i}cius and Verma, Varun and Mirin, Richard and Nam, Sae Woo and Liu, Liu and Srinivasan, Kartik},
  journal={Nature Communications},
  volume={8},
  number={1},
  pages={889},
  year={2017},
  publisher={Nature Publishing Group UK London}
}

@article{takada2025alignment,
  title={Alignment-tolerant hybrid integration of a diamond quantum sensor on a silicon nitride photonic waveguide},
  author={Takada, Kosuke and Katsumi, Ryota and Kawai, Kenta and Sato, Daichi and Yatsui, Takashi},
  journal={Optics Express},
  volume={33},
  number={11},
  pages={22769--22779},
  year={2025},
  publisher={Optica Publishing Group}
}

@article{li2024heterogeneous,
  title={Heterogeneous integration of spin--photon interfaces with a CMOS platform},
  author={Li, Linsen and Santis, Lorenzo De and Harris, Isaac BW and Chen, Kevin C and Gao, Yihuai and Christen, Ian and Choi, Hyeongrak and Trusheim, Matthew and Song, Yixuan and Errando-Herranz, Carlos and others},
  journal={Nature},
  volume={630},
  number={8015},
  pages={70--76},
  year={2024},
  publisher={Nature Publishing Group UK London}
}

@article{wan2020large,
  title={Large-scale integration of artificial atoms in hybrid photonic circuits},
  author={Wan, Noel H and Lu, Tsung-Ju and Chen, Kevin C and Walsh, Michael P and Trusheim, Matthew E and De Santis, Lorenzo and Bersin, Eric A and Harris, Isaac B and Mouradian, Sara L and Christen, Ian R and others},
  journal={Nature},
  volume={583},
  number={7815},
  pages={226--231},
  year={2020},
  publisher={Nature Publishing Group UK London}
}

@article{kim2017hybrid,
  title={Hybrid integration of solid-state quantum emitters on a silicon photonic chip},
  author={Kim, Je-Hyung and Aghaeimeibodi, Shahriar and Richardson, Christopher JK and Leavitt, Richard P and Englund, Dirk and Waks, Edo},
  journal={Nano Letters},
  volume={17},
  number={12},
  pages={7394--7400},
  year={2017},
  publisher={ACS Publications}
}

@article{mouradian2015scalable,
  title={Scalable integration of long-lived quantum memories into a photonic circuit},
  author={Mouradian, Sara L and Schr{\"o}der, Tim and Poitras, Carl B and Li, Luozhou and Goldstein, Jordan and Chen, Edward H and Walsh, Michael and Cardenas, Jaime and Markham, Matthew L and Twitchen, Daniel J and others},
  journal={Physical Review X},
  volume={5},
  number={3},
  pages={031009},
  year={2015},
  publisher={APS}
}

@article{chanana2022ultra,
  title={Ultra-low loss quantum photonic circuits integrated with single quantum emitters},
  author={Chanana, Ashish and Larocque, Hugo and Moreira, Renan and Carolan, Jacques and Guha, Biswarup and Melo, Emerson G and Anant, Vikas and Song, Jindong and Englund, Dirk and Blumenthal, Daniel J and others},
  journal={Nature Communications},
  volume={13},
  number={1},
  pages={7693},
  year={2022},
  publisher={Nature Publishing Group UK London}
}

@article{lettner2024controlling,
  title={Controlling all degrees of freedom of the optical coupling in hybrid quantum photonics},
  author={Lettner, Niklas and Antoniuk, Lukas and Ovvyan, Anna P and Gehring, Helge and Wendland, Daniel and Agafonov, Viatcheslav N and Pernice, Wolfram HP and Kubanek, Alexander},
  journal={ACS Photonics},
  volume={11},
  number={2},
  pages={696--702},
  year={2024},
  publisher={ACS Publications}
}

@article{bayer2023optical,
  title={Optical driving, spin initialization and readout of single SiV- centers in a Fabry-Perot resonator},
  author={Bayer, Gregor and Berghaus, Robert and Sachero, Selene and Filipovski, Andrea B and Antoniuk, Lukas and Lettner, Niklas and Waltrich, Richard and Klotz, Marco and Maier, Patrick and Agafonov, Viatcheslav and others},
  journal={Communications Physics},
  volume={6},
  number={1},
  pages={300},
  year={2023},
  publisher={Nature Publishing Group UK London}
}

@article{sahoo2022hybrid,
  title={Hybrid quantum nanophotonic devices with color centers in nanodiamonds},
  author={Sahoo, Swetapadma and Davydov, Valery A and Agafonov, Viatcheslav N and Bogdanov, Simeon I},
  journal={Optical Materials Express},
  volume={13},
  number={1},
  pages={191--217},
  year={2022},
  publisher={Optica Publishing Group}
}

@article{ngan2023quantum,
  title={Quantum photonic circuits integrated with color centers in designer nanodiamonds},
  author={Ngan, Kinfung and Zhan, Yuan and Dory, Constantin and Vuckovic, Jelena and Sun, Shuo},
  journal={Nano Letters},
  volume={23},
  number={20},
  pages={9360--9366},
  year={2023},
  publisher={ACS Publications}
}

@article{nader2025heterogeneous,
  title={Heterogeneous tantala photonic integrated circuits for sub-micron wavelength applications},
  author={Nader, Nima and Stanton, Eric J and Brodnik, Grant M and Jahan, Nusrat and Weight, Skyler C and Williams, Lindell M and Eshaghian Dorche, Ali and Silverman, Kevin L and Nam, Sae Woo and Papp, Scott B and others},
  journal={Optica},
  volume={12},
  number={5},
  pages={585--593},
  year={2025},
  publisher={Optica Publishing Group}
}

@article{xiang2021high,
  title={High-performance lasers for fully integrated silicon nitride photonics},
  author={Xiang, Chao and Guo, Joel and Jin, Warren and Wu, Lue and Peters, Jonathan and Xie, Weiqiang and Chang, Lin and Shen, Boqiang and Wang, Heming and Yang, Qi-Fan and others},
  journal={Nature Communications},
  volume={12},
  number={1},
  pages={6650},
  year={2021},
  publisher={Nature Publishing Group UK London}
}

@article{liang2010recent,
  title={Recent progress in lasers on silicon},
  author={Liang, Di and Bowers, John E},
  journal={Nature Photonics},
  volume={4},
  number={8},
  pages={511--517},
  year={2010},
  publisher={Nature Publishing Group UK London}
}

@article{kim2020hybrid,
  title={Hybrid integration methods for on-chip quantum photonics},
  author={Kim, Je-Hyung and Aghaeimeibodi, Shahriar and Carolan, Jacques and Englund, Dirk and Waks, Edo},
  journal={Optica},
  volume={7},
  number={4},
  pages={291--308},
  year={2020},
  publisher={Optical Society of America}
}

@article{elshaari2020hybrid,
  title={Hybrid integrated quantum photonic circuits},
  author={Elshaari, Ali W and Pernice, Wolfram and Srinivasan, Kartik and Benson, Oliver and Zwiller, Val},
  journal={Nature Photonics},
  volume={14},
  number={5},
  pages={285--298},
  year={2020},
  publisher={Nature Publishing Group UK London}
}

@article{zhu2021integrated,
  title={Integrated photonics on thin-film lithium niobate},
  author={Zhu, Di and Shao, Linbo and Yu, Mengjie and Cheng, Rebecca and Desiatov, Boris and Xin, C\_J and Hu, Yaowen and Holzgrafe, Jeffrey and Ghosh, Soumya and Shams-Ansari, Amirhassan and others},
  journal={Advances in Optics and Photonics},
  volume={13},
  number={2},
  pages={242--352},
  year={2021},
  publisher={Optical Society of America}
}

@article{xiang2022silicon,
  title={Silicon nitride passive and active photonic integrated circuits: trends and prospects},
  author={Xiang, Chao and Jin, Warren and Bowers, John E},
  journal={Photonics Research},
  volume={10},
  number={6},
  pages={A82--A96},
  year={2022},
  publisher={Chinese Laser Press and Optica Publishing Group}
}

@article{siew2021review,
  title={Review of silicon photonics technology and platform development},
  author={Siew, Shawn Yohanes and Li, Bo and Gao, Feng and Zheng, Hai Yang and Zhang, Wenle and Guo, Pengfei and Xie, Shawn Wu and Song, Apu and Dong, Bin and Luo, Lian Wee and others},
  journal={Journal of Lightwave Technology},
  volume={39},
  number={13},
  pages={4374--4389},
  year={2021},
  publisher={OSA}
}

@article{ding2024high,
  title={High-Q cavity interface for color centers in thin film diamond},
  author={Ding, Sophie W and Haas, Michael and Guo, Xinghan and Kuruma, Kazuhiro and Jin, Chang and Li, Zixi and Awschalom, David D and Delegan, Nazar and Heremans, F Joseph and High, Alexander A and others},
  journal={Nature Communications},
  volume={15},
  number={1},
  pages={6358},
  year={2024},
  publisher={Nature Publishing Group UK London}
}

@article{wan2018two,
  title={Two-dimensional photonic crystal slab nanocavities on bulk single-crystal diamond},
  author={Wan, Noel H and Mouradian, Sara and Englund, Dirk},
  journal={Applied Physics Letters},
  volume={112},
  number={14},
  year={2018},
  publisher={AIP Publishing}
}

@article{khanaliloo2015high,
  title={High-Q/V monolithic diamond microdisks fabricated with quasi-isotropic etching},
  author={Khanaliloo, Behzad and Mitchell, Matthew and Hryciw, Aaron C and Barclay, Paul E},
  journal={Nano Letters},
  volume={15},
  number={8},
  pages={5131--5136},
  year={2015},
  publisher={ACS Publications}
}

@article{burek2012free,
  title={Free-standing mechanical and photonic nanostructures in single-crystal diamond},
  author={Burek, Michael J and De Leon, Nathalie P and Shields, Brendan J and Hausmann, Birgit JM and Chu, Yiwen and Quan, Qimin and Zibrov, Alexander S and Park, Hongkun and Lukin, Mikhail D and Loncar, Marko},
  journal={Nano Letters},
  volume={12},
  number={12},
  pages={6084--6089},
  year={2012},
  publisher={ACS Publications}
}

@article{zhan2020deterministic,
  title={Deterministic generation of loss-tolerant photonic cluster states with a single quantum emitter},
  author={Zhan, Yuan and Sun, Shuo},
  journal={Physical Review Letters},
  volume={125},
  number={22},
  pages={223601},
  year={2020},
  publisher={APS}
}

@article{buterakos2017deterministic,
  title={Deterministic generation of all-photonic quantum repeaters from solid-state emitters},
  author={Buterakos, Donovan and Barnes, Edwin and Economou, Sophia E},
  journal={Physical Review X},
  volume={7},
  number={4},
  pages={041023},
  year={2017},
  publisher={APS}
}

@article{nguyen2019quantum,
  title={Quantum network nodes based on diamond qubits with an efficient nanophotonic interface},
  author={Nguyen, CT and Sukachev, DD and Bhaskar, MK and Machielse, Bartholomeus and Levonian, DS and Knall, EN and Stroganov, Pavel and Riedinger, Ralf and Park, Hongkun and Lon{\v{c}}ar, M and others},
  journal={Physical Review Letters},
  volume={123},
  number={18},
  pages={183602},
  year={2019},
  publisher={APS}
}

@article{bhaskar2020experimental,
  title={Experimental demonstration of memory-enhanced quantum communication},
  author={Bhaskar, Mihir K and Riedinger, Ralf and Machielse, Bartholomeus and Levonian, David S and Nguyen, Christian T and Knall, Erik N and Park, Hongkun and Englund, Dirk and Lon{\v{c}}ar, Marko and Sukachev, Denis D and others},
  journal={Nature},
  volume={580},
  number={7801},
  pages={60--64},
  year={2020},
  publisher={Nature Publishing Group UK London}
}

@article{sun2018single,
  title={A single-photon switch and transistor enabled by a solid-state quantum memory},
  author={Sun, Shuo and Kim, Hyochul and Luo, Zhouchen and Solomon, Glenn S and Waks, Edo},
  journal={Science},
  volume={361},
  number={6397},
  pages={57--60},
  year={2018},
  publisher={American Association for the Advancement of Science}
}

@article{huber2018semiconductor,
  title={Semiconductor quantum dots as an ideal source of polarization-entangled photon pairs on-demand: a review},
  author={Huber, Daniel and Reindl, Marcus and Aberl, Johannes and Rastelli, Armando and Trotta, Rinaldo},
  journal={Journal of Optics},
  volume={20},
  number={7},
  pages={073002},
  year={2018},
  publisher={IOP Publishing}
}

@article{aharonovich2016solid,
  title={Solid-state single-photon emitters},
  author={Aharonovich, Igor and Englund, Dirk and Toth, Milos},
  journal={Nature Photonics},
  volume={10},
  number={10},
  pages={631--641},
  year={2016},
  publisher={Nature Publishing Group UK London}
}

@article{senellart2017high,
  title={High-performance semiconductor quantum-dot single-photon sources},
  author={Senellart, Pascale and Solomon, Glenn and White, Andrew},
  journal={Nature Nanotechnology},
  volume={12},
  number={11},
  pages={1026--1039},
  year={2017},
  publisher={Nature Publishing Group UK London}
}

@article{toninelli2021single,
  title={Single organic molecules for photonic quantum technologies},
  author={Toninelli, C and Gerhardt, I and Clark, AS and Reserbat-Plantey, A and G{\"o}tzinger, Stephan and Ristanovi{\'c}, Z and Colautti, M and Lombardi, P and Major, KD and Deperasi{\'n}ska, I and others},
  journal={Nature Materials},
  volume={20},
  number={12},
  pages={1615--1628},
  year={2021},
  publisher={Nature Publishing Group UK London}
}

@article{bradac2019quantum,
  title={Quantum nanophotonics with group IV defects in diamond},
  author={Bradac, Carlo and Gao, Weibo and Forneris, Jacopo and Trusheim, Matthew E and Aharonovich, Igor},
  journal={Nature Communications},
  volume={10},
  number={1},
  pages={5625},
  year={2019},
  publisher={Nature Publishing Group UK London}
}

@article{janitz2020cavity,
  title={Cavity quantum electrodynamics with color centers in diamond},
  author={Janitz, Erika and Bhaskar, Mihir K and Childress, Lilian},
  journal={Optica},
  volume={7},
  number={10},
  pages={1232--1252},
  year={2020},
  publisher={Optical Society of America}
}

@article{atature2018material,
  title={Material platforms for spin-based photonic quantum technologies},
  author={Atature, Mete and Englund, Dirk and Vamivakas, Nick and Lee, Sang-Yun and Wrachtrup, Joerg},
  journal={Nature Reviews Materials},
  volume={3},
  number={5},
  pages={38--51},
  year={2018},
  publisher={Nature Publishing Group UK London}
}

@article{chang2014quantum,
  title={Quantum nonlinear optics—photon by photon},
  author={Chang, Darrick E and Vuleti{\'c}, Vladan and Lukin, Mikhail D},
  journal={Nature Photonics},
  volume={8},
  number={9},
  pages={685--694},
  year={2014},
  publisher={Nature Publishing Group UK London}
}

@article{azuma2023quantum,
  title={Quantum repeaters: From quantum networks to the quantum internet},
  author={Azuma, Koji and Economou, Sophia E and Elkouss, David and Hilaire, Paul and Jiang, Liang and Lo, Hoi-Kwong and Tzitrin, Ilan},
  journal={Reviews of Modern Physics},
  volume={95},
  number={4},
  pages={045006},
  year={2023},
  publisher={APS}
}

@article{basani2025universal,
  title={Universal logical quantum photonic neural network processor via cavity-assisted interactions},
  author={Basani, Jasvith Raj and Niu, Murphy Yuezhen and Waks, Edo},
  journal={npj Quantum Information},
  volume={11},
  number={1},
  pages={142},
  year={2025},
  publisher={Nature Publishing Group UK London}
}

@article{kok2007linear,
  title={Linear optical quantum computing with photonic qubits},
  author={Kok, Pieter and Munro, William J and Nemoto, Kae and Ralph, Timothy C and Dowling, Jonathan P and Milburn, Gerard J},
  journal={Reviews of Modern Physics},
  volume={79},
  number={1},
  pages={135--174},
  year={2007},
  publisher={APS}
}

@article{wang2020integrated,
  title={Integrated photonic quantum technologies},
  author={Wang, Jianwei and Sciarrino, Fabio and Laing, Anthony and Thompson, Mark G},
  journal={Nature Photonics},
  volume={14},
  number={5},
  pages={273--284},
  year={2020},
  publisher={Nature Publishing Group UK London}
}

@article{bopp2024sawfish,
  title={‘Sawfish’Photonic Crystal Cavity for Near-Unity Emitter-to-Fiber Interfacing in Quantum Network Applications},
  author={Bopp, Julian M and Plock, Matthias and Turan, Tim and Pieplow, Gregor and Burger, Sven and Schr{\"o}der, Tim},
  journal={Advanced Optical Materials},
  volume={12},
  number={13},
  pages={2301286},
  year={2024},
  publisher={Wiley Online Library}
}

@misc{Tidy3D,
  author       = {{Flexcompute, Inc.}},
  title        = {Tidy3D, next-gen electromagnetic simulation tool},
  howpublished = {\url{https://www.flexcompute.com/tidy3d/solver/}},
  year         = {2024},
}

\end{document}


\setcounter{figure}{0}
\renewcommand{\figurename}{Fig.}
\renewcommand{\thefigure}{S\arabic{figure}}
\setcounter{secnumdepth}{1}

\setcounter{section}{0}
\renewcommand{\thesection}{S\arabic{section}}

\title{Self‑Aligned Heterogeneous Quantum Photonic Integration}
\author{Kinfung Ngan}
\affiliation{
JILA and Department of Physics, University of Colorado Boulder, Colorado 80309, USA
}
\author{Yeeun Choi}
\affiliation{
Center for Quantum Technology, Korean Institute of Science and Technology (KIST), Seoul 02792, Republic of Korea
}
\affiliation{KU-KIST Graduate School of Converging Science and Technology, Korea University, Seoul, Republic of Korea}
\author{Chun-Chieh Chang}
\affiliation{
Center for Integrated Nanotechnologies, Los Alamos National Laboratory, Los Alamos, New Mexico 87545, USA
}
\author{Dongyeon Daniel Kang}
\email{dykang@kist.re.kr}
\affiliation{
Center for Quantum Technology, Korean Institute of Science and Technology (KIST), Seoul 02792, Republic of Korea
}
\affiliation{
Division of Quantum Information, KIST School, Korea University of Science and Technology (UST), Seoul 02792, Republic of Korea
}
\author{Shuo Sun}
\email{shuosun@colorado.edu}
\affiliation{
JILA and Department of Physics, University of Colorado Boulder, Colorado 80309, USA
}

\date{\today}

\maketitle

\section{inverse-designed grating coupler}

\begin{figure*}[htp]
     \centering
         \includegraphics[width=0.8\textwidth]{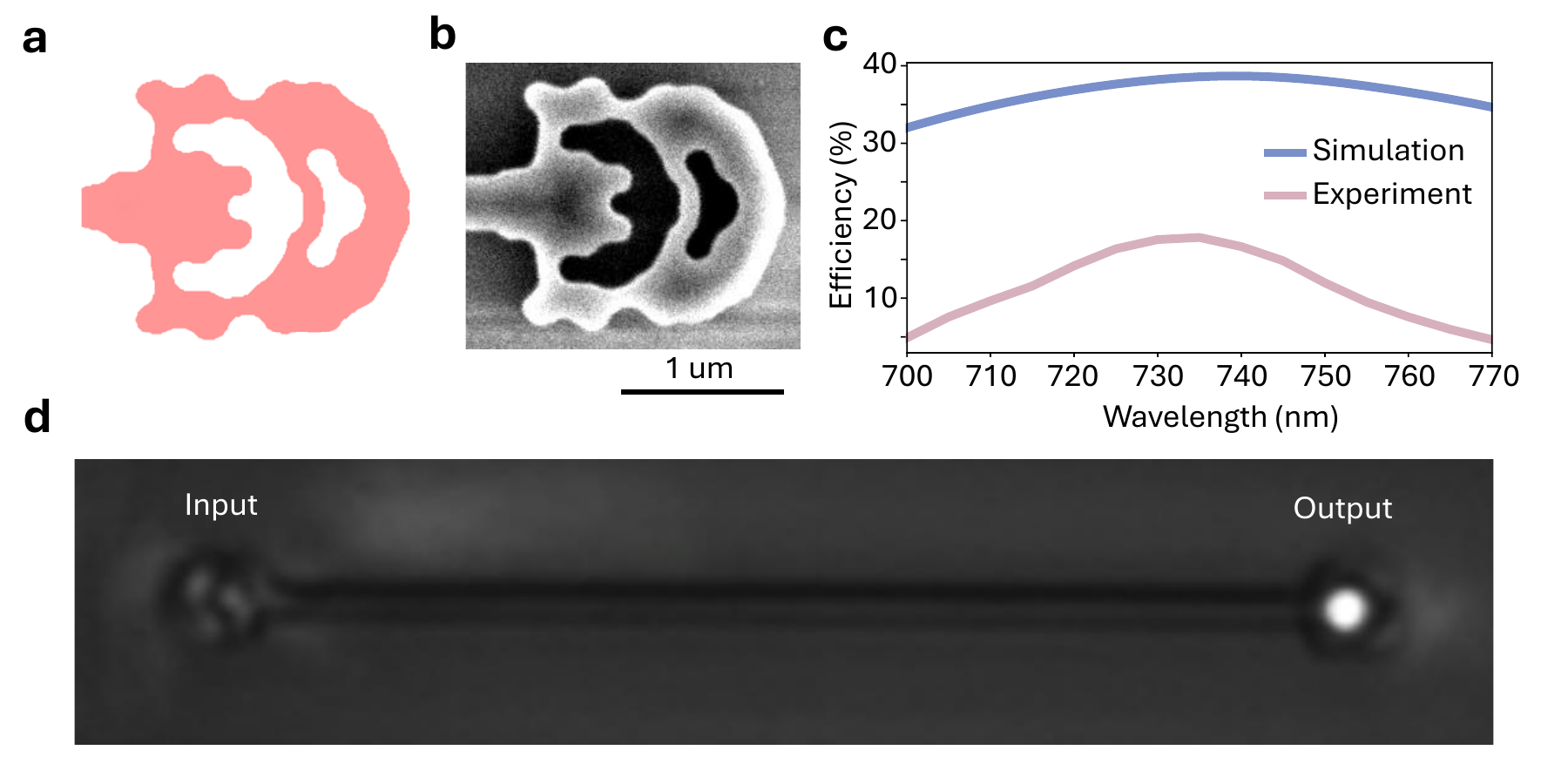}
         \caption{Design and characterization of the inverse-designed grating coupler. (a) Design of the vertical grating coupler. (b) SEM image of the fabricated grating coupler. (c) The coupling efficiency of the grating coupler as a function of wavelength. Blue solid line shows the calculated values, and red solid line shows the experimentally measured values. (d) Optical microscope image of a waveguide when excited by a laser at the input grating coupler. A clear Gaussian-like emission at the output is visible, indicating good coupling efficiency from the Gaussian beam to the waveguide fundamental mode.}
         \label{S1}
\end{figure*}

Figure~\ref{S1}a shows the design of the vertical grating coupler used in this work. The inverse-designed structure was obtained via adjoint-based topology optimization implemented with Tidy3D (Flexcompute) \color{blue} \cite{Tidy3d}\color{black}. The coupler has a footprint of 1.4$\times$1.6 $\mu\mathrm{m}^{2}$ and is optimized to couple the fundamental waveguide mode into a vertically emitted Gaussian beam with a beam waist diameter of 0.52~$\mu$m. The blue solid line in Fig.~\ref{S1}c shows the calculated coupling efficiency as a function of wavelength. The design achieves a peak efficiency of 38\%. Notably, although we did not apply multi-frequency optimization, the design naturally exhibits a broad coupling bandwidth, maintaining an efficiency exceeding 30\% over a window of more than 70~nm.

Figure~\ref{S1}b shows a scanning electron microscope (SEM) image of the inverse-designed grating coupler. To characterize the coupling efficiency, we fabricated a waveguide terminated by identical couplers at both ends. Figure~\ref{S1}d shows a confocal optical microscope image of this waveguide. When we excite the input grating coupler with a continuous-wave laser, we observe a strong Gaussian-like emission from the output coupler, indicating efficient coupling. The red solid line in Fig.~\ref{S1}c shows the experimentally measured coupling efficiency of a single grating coupler as the laser wavelength is tuned. We achieve a peak efficiency of 18\% with a full width at half maximum (FWHM) of 40~nm. We note that this value is a lower bound because our calibration assumes perfect coupling between the Gaussian mode and the fiber, and neglects waveguide propagation loss.

\section{Limitations of the cavity quality factor}
\begin{figure*}[htp]
     \centering
         \includegraphics[width=0.9\textwidth]{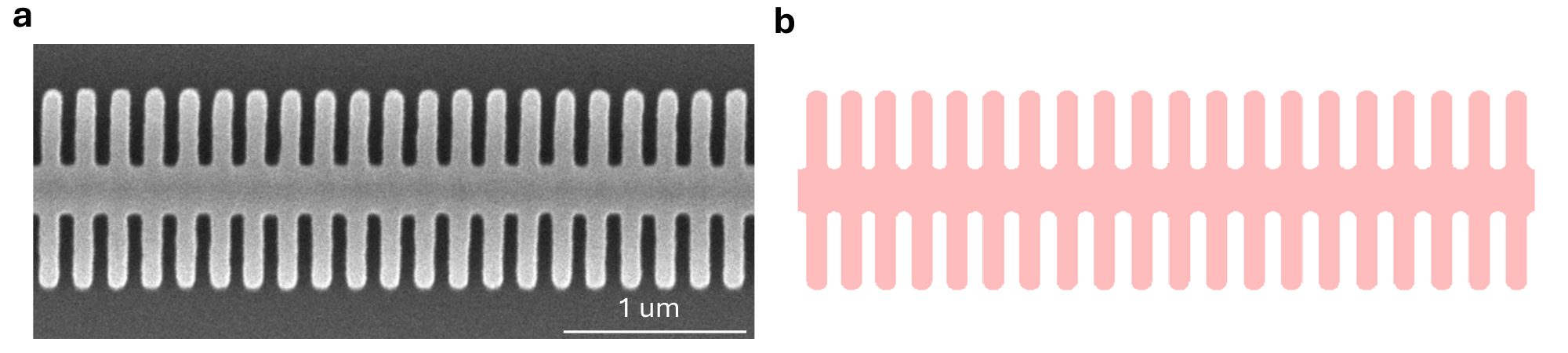}
         \caption{SEM image (a) and the extracted pattern (b) of the fabricated fishbone photonic crystal cavity.
}
         \label{S2}
\end{figure*}

In the main text, we report a measured cavity $Q$ factor of 4,600, which is 20 times lower than the simulated value. To investigate the origin of the reduced $Q$, we took an SEM image of the same cavity (Fig.~\ref{S2}a) and recalculated the cavity $Q$ factor using the cavity geometry extracted from the SEM image (Fig.~\ref{S2}b). The calculated $Q$ factor of the extracted cavity geometry is 5,200, with a resonance near 730~nm, in good agreement with our experimental measurement. We therefore conclude that the cavity $Q$ factor is primarily limited by geometric deviations between the design and the fabricated device, rather than by scattering or absorption introduced by the heterogeneous integration or other steps in our fabrication process.

To identify which geometric deviations contribute most to the $Q$ reduction, we performed additional simulations starting from the designed geometry. Rounding all corners to the same extent as in the fabricated device reduces the calculated $Q$ from 120,000 to 30,000. Incorporating the observed fin-width variations further reduces the $Q$ from 30,000 to 5,200. These results suggest a clear path toward improving the cavity $Q$ through better calibrated electron beam exposure, including optimizing the exposure dose and applying appropriate proximity effect correction. These steps should increase the cavity $Q$ by maintaining uniform fin widths and minimizing corner rounding. More broadly, our findings indicate that alternative photonic crystal designs that intentionally incorporate rounded-corner features~\cite{bopp2024sawfish} may offer improved fabrication tolerance compared to our original design. We will explore both fabrication improvements and more fabrication-tolerant designs in future work. Once the measured cavity $Q$ is no longer limited by electron beam exposure, we can then assess any additional limitations imposed by the heterogeneous integration process.

\section{Characterization and tuning of the cavity used for Purcell enhancement}

The photonic crystal cavity used for the Purcell-enhancement demonstration (Figs.~3d and 3e of the main text) was fabricated on a different chip from the cavity shown in Fig.~3c of the main text. This chip was fabricated when the electron beam lithography system was suffering from large electron source instability. For these reasons, cavities on this chip exhibit significantly lower $Q$ factors.

\begin{figure*}[htp]
     \centering
         \includegraphics[width=\textwidth]{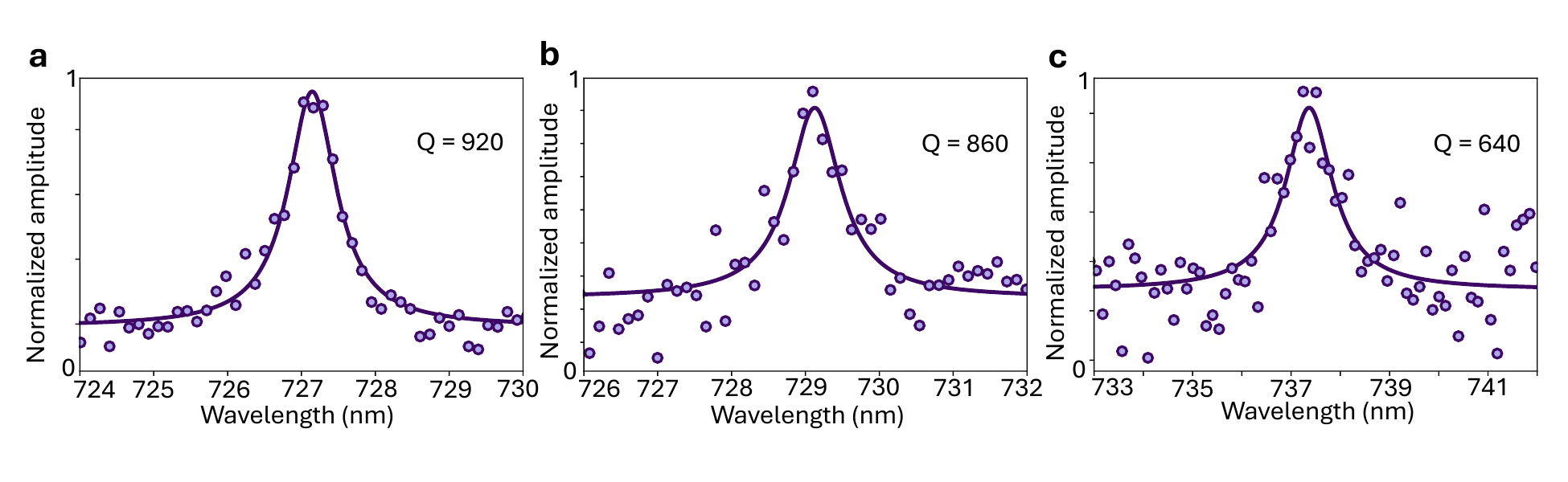}
         \caption{Spectrum of the photonic crystal cavity used for the Purcell enhancement experiment (Fig.~3d and 3e of the main text) right after the device fabrication (a), after 20 ALD deposition cycles (b), and after a total of 120 ALD deposition cycles (c). The cavity spectrum was measured from a weak and broadband TiO$_2$ autofluorescence that is enhanced at the cavity resonance. In all panels, the circles are the measured data, and the solid line is a fit to a Lorentzian function from which we extract the cavity $Q$ factor.}
         \label{S3}
\end{figure*}

Figure~\ref{S3}a shows the spectrum of the photonic crystal cavity used for the Purcell-enhancement experiment. The cavity $Q$ factor is 920, which is only $\sim1/5$ of that reported of the cavity in Fig.~3c of the main text. In addition, the cavity resonance is at $\sim$727~nm, which is blue-detuned from the SiV zero-phonon line by 10~nm. This detuning exceeds the in-situ tuning range achievable via gas condensation ($\sim$4~nm). We therefore explored an alternative strategy for cavity wavelength tuning. By depositing TiO$_2$ on top of the device via atomic layer deposition (ALD), we can red shift the cavity resonance, and the exact amount of shift depends on the ALD layer thickness. Figure~\ref{S3}b shows the measured cavity spectrum after 20 ALD cycles. We observe a shift of $\sim2$~nm and, at the same time, a reduction in cavity $Q$ from 920 to 860, which we attribute to a mismatch in cavity thickness between the design and the post-deposition device. Figure~\ref{S3}c shows the measured cavity spectrum after 120 ALD cycles. We successfully shift the cavity resonance to 737~nm. However, the cavity $Q$ decreases further to 640. This reduced $Q$ limits the Purcell factor achievable with this device.

\section*{References}

\bibliography{ref_supplement}